\documentclass[prl,twocolumn]{revtex4}
\usepackage{epsfig}
\begin{document}
\title[Field-induced metal-insulator]{Field-induced metal-insulator 
transition in the $c-$axis resistivity of graphite} 
\date{\today}
\author{H. Kempa}
\author{P. Esquinazi}
\email[E-mail: ]{esquin@physik.uni-leipzig.de}
\affiliation{Abteilung  Supraleitung und
Magnetismus, Institut f\"ur Experimentelle Physik II, 
Universit\"at Leipzig, Linn{\'e}str. 5, D-04103 Leipzig,
Germany}  
\author{Y. Kopelevich} 
\affiliation{Instituto de F{\'i}sica, 
Universidade Estadual de Campinas, 
Unicamp 13083-970, Campinas, S\~{a}o Paulo, Brasil}

\begin{abstract}
We show that the resistivity perpendicular $\rho_c$ and parallel $\rho_a$ 
to the basal planes of different graphite samples show 
similar magnetic-field-driven metal-insulator-transitions at a field $B_c \sim 0.1~$T applied parallel to the 
$c-$axis. Our results demonstrate the universality of the recently found scaling in $\rho_a$  
of  graphite and  indicate that the metallic-like
temperature dependence of  $\rho_c$ is directly correlated to that of $\rho_a$. The similar magnetoresistance 
found for both resistivities, the violation of Kohler's rule
and the field-induced transition indicate that the semiclassical transport theory is inadecuate to understand
the transport properties of graphite.
\end{abstract}

\maketitle

Although a considerable amount of studies has been performed on
graphite and related compounds, their transport properties are still not well
understood. The scientific interest on graphite has been recently renewed by 
 new magnetization \cite{kopejltpc} and transport \cite{kopessc} results 
on highly oriented pyrolytic graphite (HOPG). These 
show irreversible magnetization \cite{kopejltpc} that, upon sample, its previous 
thermal treatment and magnetic field orientation, resembles that of a 
superconducting loop even at room temperature 
suggesting the existence of localized superconducting domains 
at high temperatures in topological disordered regions \cite{gon}. This result
is indirectly supported by the recently found magnetic-field-driven superconductor-insulator-type
transition (SIT) in the in-plane resistivity of a HOPG sample \cite{kopessc}. 
Remarkably, this field-driven transition 
showed a similar scaling as found
for two-dimensional (2D) disordered superconductors as well as for Si MOSFETs \cite{review}. 

It has been speculated that in graphite 
impurity-assisted hopping 
mechanism may affect the electronic transport along the $c-$axis. In this case 
the conduction across the graphite planes might be influenced by the in-plane transport
in a given temperature $(T)$ range. The aim of our research is to check whether  the $c-$axis resistivity $\rho_c(T,B)$ 
 shows a similar SIT with a similar scaling as the in-plane resistivity $\rho_a(T,B)$. Evidence for this 
speculation has  important consequences to solve
 an old and basic problem of the transport in graphite, namely, to what extent the metallic-like
behavior of $\rho_c$ is an intrinsic property of the transport along the $c-$axis and
whether the semiclassical theories are appropriate to account for the observed effects.
Furthermore, it is important to prove experimentally the universality of the 2D scaling found
in graphite; for this aim different graphite samples with different resistivities have been measured
in this work.

We have studied the resistivities 
$\rho_a(T,B)$ and  $\rho_c(T,B)$ of four  samples: two HOPG samples,  
sample 1 from Advanced Ceramics (rocking curve FWHM $\simeq 0.4^o$), sample 2 (from the Research 
Institute ``Graphite" (Moscow),  
FWHM $\simeq 0.4^o$) and samples 3 and 4,  Kish graphite crystals. The samples typical 
length and width were $\sim 2~$mm and thickness between 0.1~mm and 0.5~mm. 
 The in-plane resistivity  $\rho_a$ was measured attaching four contacts on the
 sample surface with silver paint. The absolute values were obtained with other configuration that assures 
a uniform current distribution. The $c-$axis resistance was measured attaching two contacts on 
each of the surfaces (one in the middle and other surroundig it and contacted on the rest surface). A 
resistance bridge (LR700) was used with an ac current $\le 1~$mA. The field was
applied normal to the graphene layers.  The
temperature stability was better than 2 mK in the whole range. 

The left  panels of Figs. 1 to 3 show both resistivities as a function of $T$ at constant fields for samples
1 to 3, respectively.  At zero applied field the basal-plane resistivity shows the typical behavior reported
in the literature: whereas for sample 1 $\rho_a$ shows a metallic behavior up to $T \sim 100~$K,  
 ($\sim 220$~K for sample 2), sample 3 -- which shows  the lowest resistivity -- 
up to the highest measured $T \sim 270~$K. 
 Similar  absolute values as well as the
 observed maximum of $\rho_c$ for samples 1 and 3 were reported in the 
literature  \cite{tsu,matsu}. Note also
that sample 2 with the highest $\rho_c(0)$ shows no maximum but a saturation below $\sim 10~$K. 
The  basal-plane rest resistance ratio $RRR = R(300K)/R(4.2)$ has been
tentatively taken in the literature as a measure of sample perfection \cite{tsu}. 
A speculative interpretation of the
observed  non-monotonic behavior of $\rho_c$ vs. $RRR$ has been given in terms
of the influence of laminar defects and microcraks. Samples with $RRR < 10$, for example, were
considered of poor perfection. We note however, that this sample classification
might be misleading because topological defects
may play a role other than scattering centers. It has been recently demonstrated that defects can lead to a local
enhancement of the density of states and generate superconducting as well as ferromagnetic
domains \cite{gon}. In this case large $RRR$ values do not necessarily mean high quality. We believe that a broad
spectrum of transport data of graphite may need to be reinterpreted.

A clear magnetic field-driven transition from a metallic- to a semiconducting-like behavior  is 
observed in both $\rho_a$ and
$\rho_c$, see Figs. 1 to 3. Remarkably, this transition is observed at fields $B \sim 0.1~$T in both  
resistivities. The observed behavior in $\rho_a(T,B)$, in particular the leveling at low $T$, is similar to that reported 
in 2D films \cite{mason}. The minimum in $\rho_a(T)$, that develops at intermediate fields
in sample 3, is similar to that measured in a Si MOSFET \cite{simo} but  located
at 100 times higher $T$ and observed at 10 times smaller fields \cite{epr}. In spite of these similarities,
the transition in graphite may have other origin as in Si MOSFETs. We note that the 
metal-insulator transition (MIT) in 2D electron systems 
is measured in the in-plane resistivity as a function of the magnetic field applied parallel as well as 
normal to the film plane \cite{review}. Recent high-resolution angle-dependent experiments \cite{kempa} 
reveal however, that the MIT in graphite is driven mainly by the perpendicular component of the 
field providing a clear evidence of the 2D character of graphite in agreement with recent theoretical 
work \cite{khv}.  

In the ideal case the separatrix line obtained 
at a critical field $B = B_c$  is a line where the 
resistivity  remains temperature independent and separates the metallic from the semiconducting branch. 
It has been shown that for 2D disordered superconductors 
as well as for Si MOSFET's, for fields near the MIT  the
 in-plane resistivity follows a scaling when
it is plotted as function of the scaling variable $|B- B_c| T^{-1/\alpha}$, being  $\alpha$ a critical
exponent. In a similar way we determined $B_c$ as well as $\alpha$ from
$\rho_a(T,B)$, see the right upper panels in Figs.~1 to 3. Taking into account 
that the separatrix line is not an ideal 
horizontal line in our samples, we obtain a reasonable scaling in a restricted temperature range. 
 Both parameters are similar for the three samples and agree with those
reported in \cite{kopessc}. The exponent $\alpha =0.6 \pm 0.05$ coincides 
with that  for Si MOSFETs \cite{kra} and ultrathin a-Bi films \cite{yaz}. A similar scaling with the same 
exponent $\alpha$ but with 
a smaller $B_c$ is obtained for $\rho_c$ at 30~K~$\le T \le 60~$K for samples 1 and 3.
Sample 2 $c-$axis resistivity shows no metallic behavior at zero field and therefore no
apparent scaling. Based on this scaling a SIT in MOSFETs has been suggested \cite{phi}.
The evidence for Cooper pairing of the carriers, however,  is still a matter of discussion in the
literature and no consent has been achieved yet \cite{review}.

As shown in superconducting amorphous In-O films
\cite{gant} the scaling can be considerably improved if one assumes
that in addition to the contribution to the conductivity that shows the 
SIT,  an additional $T-$dependent contribution
affects the resistivity. It has been speculated that this contribution can be due to
the conductivity of normal electrons \cite{gant}. Within the same idea and 
assuming that the intrinsic $c-$axis
resistivity of graphite has a non-metallic $T-$dependence which adds as background
to the field dependent part, we show in the low right 
panels of Figs. 1 to 3 the scaling obtained if we subtract from the measured $\rho_c(T,B)$
the line $\rho_c(T,B_c)$ or a linear $T-$dependence $a \rho_0 T$ (sample 2, $a$ is a free
parameter and $\rho_0 = \rho_c(T \rightarrow 0, B \sim 0.1~$T)). In this case the $T-$range for scaling  is considerable
enhanced. Obviously, a similar inprovement of the $T-$range for scaling is obtained also for the
in-plane resistivity after subtracting a $T-$dependent background. The value of  the exponent 
$\alpha$ remains the same. These results indicate the universality of the MIT found in graphite
for both resistivities and 
of its scaling.

Graphite can be considered as a dilute electron system with a Coulomb coupling constant
$r_s = 1/(\pi n_{2D})^{1/2}a^*_B$ where the 2D carrier density is given by $n_{2D} = n_{3D} d$ with
$d = 0.335~$nm the interplane distance, $a^*_B=\epsilon\hbar^2/e ^2m^*$ the effective Bohr radius,
$\epsilon = 2.8$ the dielectric constant and $m^*$ the effective mass of the carriers. 
Taking literature values for the majority (minority) carriers \cite{lit} one
obtains $r_s \sim 5 (10)$ suggesting that the 2D electron system in graphite might be considered as 
a strongly correlated liquid and 
the effects due to the electron-electron interaction 
should be important. We note that  $n_{2D} \sim 10^{11}$~cm$^{-2}$ practically
coincides with that reported for a Si MOSFET \cite{review}.

One important characteristic observed in several of the measured 2D systems 
with a MIT is that the critical sheet resistance  $R_s$ at the separatrix is of the order of the
quantum unit $h/e^2$. Within the scaling theory \cite{fischer} 
of a SIT, $R_s = h/4 e^2 \simeq 6.4~$k$\Omega$ at $B_c$. Assuming that the input current
decreases exponentially in the $c-$direction with a penetration depth $\lambda$
and taking into account the distance $d$ between graphene sheets, we
estimate $R_s \simeq (b/l) R /(1 - \exp(-d/\lambda))$ for a single graphene layer, 
where $R$ is the measured longitudinal resistance at the sample surface, $b$ the sample 
width and $l$ the distance
between voltage contacts \cite{kempadiplom}. With measured parameters we obtain at the separatrix
$R_s \simeq  a (h/4e^2)$ with $a = 1/3.2 (1/3.7)$ for sample 1 (2), whereas for sample 3
$a = 1/27$, using $\lambda \simeq 10~\mu$m obtained from 
measurements with different configurations. 
The difference in sheet resistance between samples 1,2 and 3 may be related to
a difference in the coupling between layers due to lattice defects, like stacking
faults or in-plane defects, or may indicate the lack of an unique critical
resistance as observed in disordered thin films \cite{pp}.

We note that lattice defects may not only affect the transport as scattering centers but they
may contribute to enhance the electronic coupling between
layers giving rise to a quasi 3D electronic spectrum with coherent transport along
the $c-$axis. Field-angle dependent experiments around the direction parallel to the planes show
 a weak coherent peak in the $c-$axis resistivity for Kish graphite (but not for the HOPG samples) 
as expected from theoretical considerations for coherent transport \cite{coherent}. 
Regarding defects within the graphene layers that may generate superconducting domains \cite{gon}, we
note that for sample 3 the metallic behavior of $\rho_c$ measured at low $T$ remains at all fields
irrespective of the behavior of $\rho_a$ which shows an insulating-like behavior at fields
$\ge  300~$Oe. This behavior in Kish graphite might be due to a ``phase separation", i.e. 
to the presence of both insulating and metallic domains within the planes.

The $T-$dependence of 
 $\rho_c$ in HOPG has been interpreted \cite{matsu} in terms of a semiclassical 
model with parameters like
the carrier density $n(T)$, the stacking-fault spacing $l$ and the potential
barrier formed on the plane of stacking disorder $\Delta E$. 
The increase of $\rho_c$ with $T$ at $T < 100~$K was attributed
to an interlayer transfer of carriers which
provides the mixing of the $c$-axis and in-plane transport. It was further
speculated that the interaction of the carriers with phonons on the planes 
determines the $T-$dependence \cite{matsu}. 
We note first that according to the model \cite{matsu} and from the comparison of the absolute values
of $\rho_c$ of our samples,  sample 2 should have an extremely  large stacking-fault
density. This is at odd with x-ray diffraction results that indicate that this kind of lattice 
defects is absent in high quality HOPG \cite{sf}. Second, from our results it is 
obvious that the increase of $\rho_c$ with $T$ at 
$T < 100~$K cannot be attributed simply to electron-phonon scattering. Nevertheless, the concept of 
conduction-path mixing mechanism is appealing. This 
can be realized if the conduction-path of the 
electrons along the $c-$axis is in part short circuited by lattice defects or impurities
as was usually speculated in literature. 
However, the apparent similarity of the behavior obtained in the three samples may
indicate an intrinsic origin for the conduction-path mixing. A situation similar to that
found in layered high-temperature superconductors, where
in-plane charge fluctuations influence  the $c-$axis transport \cite{tur}, may be possible in graphite.
Our results cast also doubts about the estimate of $\sim 0.3~$eV for the interlayer transfer integral used
in literature \cite{lit}. This is not surprising since neither the electron-electron interaction nor
charge fluctuations \cite{voz} were taken into account. 

Regarding the field dependence of $\rho_c$
one may argue that
the field can decrease the carrier concentration at low $T$ inducing an  increase in $\rho_a$ and $\rho_c$.
However, this appears to be at odd with the  increase of $n(T)$ by a factor of two with a field
of  0.2 T obtained from magnetoresistance measurements and within a semiclasical model  \cite{dillon}.
As noted in the literature \cite{dillon} the mean mobility in graphite cannot be estimated rigurously with the usual
two-band model since the resistivity does not follow the predicted $B^2$ dependence.  Arbitrary definitions 
to circumvent this problem
did not provide a satisfactory description of the measurements.
Our results indicate clearly that 
Kohler's rule is not fulfilled in graphite, neither in the $a$ nor in the $c-$direction   
--as can be seen in  Fig.~4(a) for $\rho_a$--  implying that
the semiclassical picture does not apply. 
Furthermore, we demonstrate in Fig.4(b) that the low-temperature 
 magnetoresistance
in the $c$-direction, shown for two Kish graphite samples and sample 1, is basically the same 
as the in-plane one. This result also indicates the correlation between both conductivities
and contradicts the semiclassical model since for $B || c || J$ no magnetoresistance
in the $c-$direction is expected.

Concluding, we proved
that both in-plane and out-of-plane resistivities in graphite show a
MIT with a scaling similar to 2D electron systems and indicate that the
metallic-like behavior of 
$\rho_c(T,B)$ is related to $\rho_a(T,B)$. Our overall results
show that semiclassical models cannot account for the magnetotransport
properties of graphite.

\begin{acknowledgments}
This work is supported by  DFG  Grant ES 86/6-1. 
The support from the DAAD, CAPES, CNPq and FAPESP is gratefully acknowledge.
We thank H.-C. Semmelhack for  fruitful discussions. 
\end{acknowledgments}

\begin{widetext}
\newpage
\begin{figure}
\centerline{\psfig{file=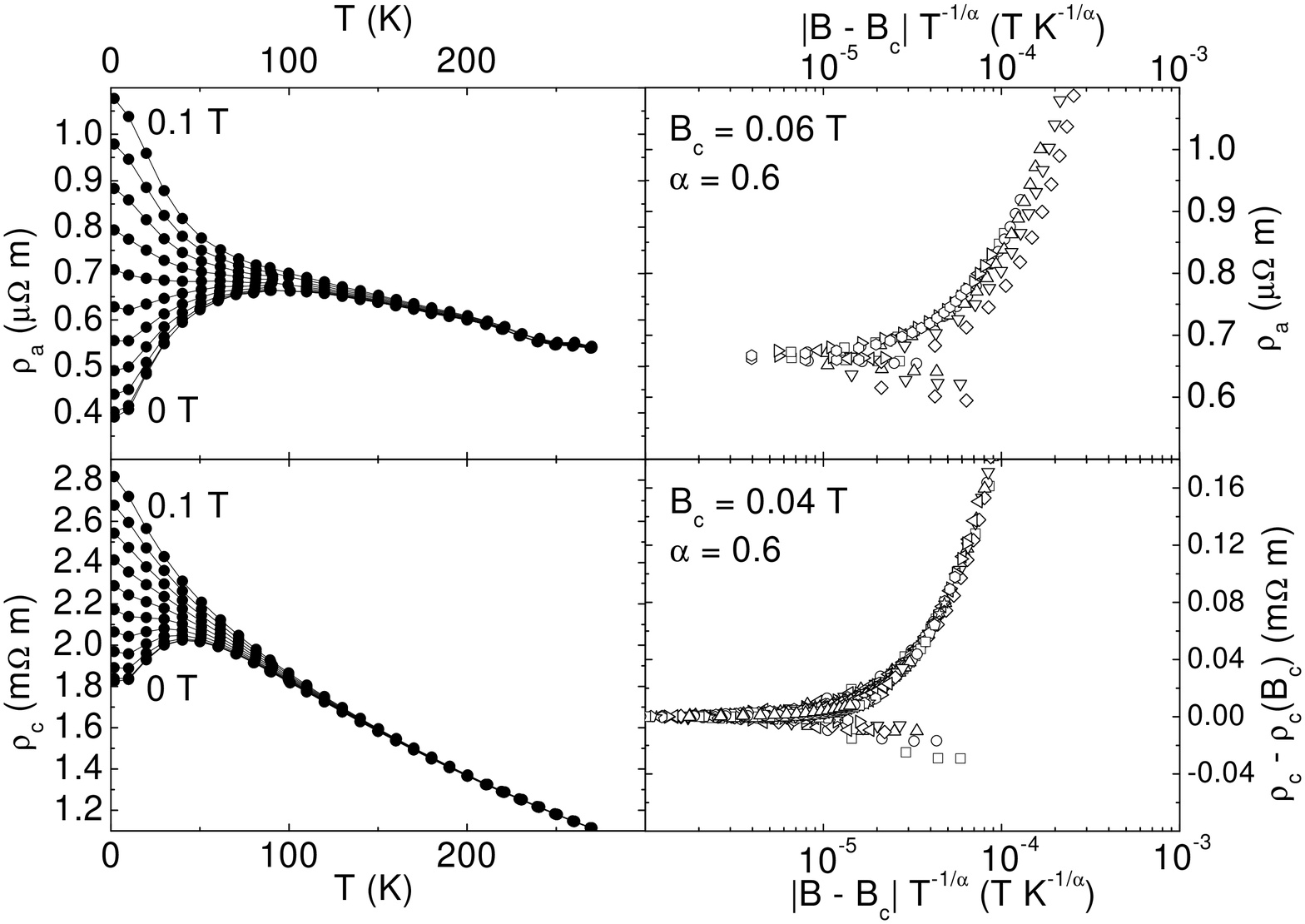,height=4.8in}}
\caption{Left panels: In-plane $\rho_a$ and $c-$axis $\rho_c$ resistivities as a function
of temperature at constant fields for sample 1. The field step between two curves is 0.01~T. 
Right panels: Scaling behavior of the resistivities. The experimental line $\rho_c(B_c= 0.04~$T)
 has been subtracted from the $\rho_c$-data. 
The symbols are the data points obtained between 30~K and 120~K. The symbols in the lower figure 
correspond to the data taken between 40~K and 270~K.}
\label{1}
\end{figure}

\newpage

\begin{figure}
\centerline{\psfig{file=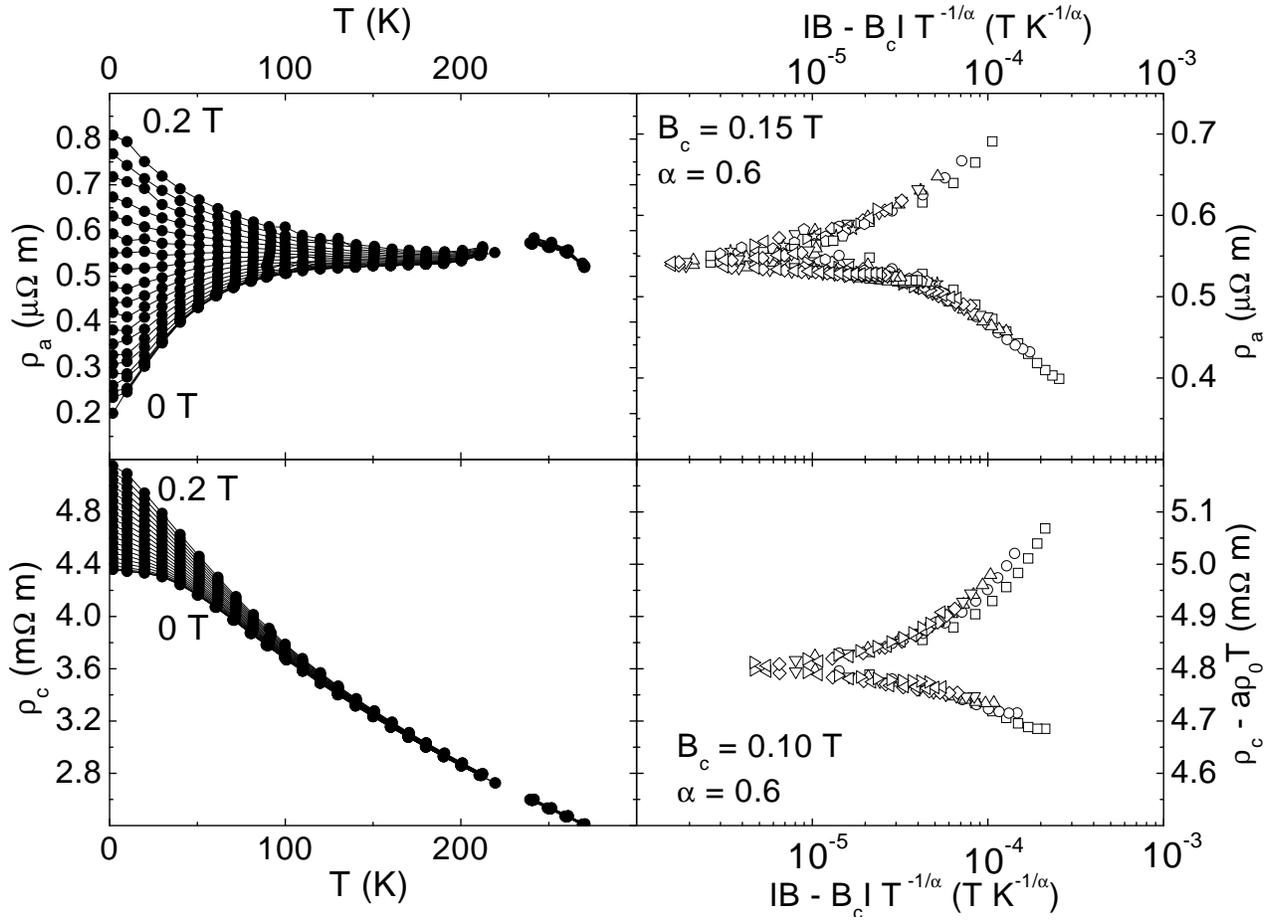,height=4.8in}}
\caption{The same as Fig.~1 but for sample 2. Right pannel: 
The symbols are the data points obtained between 30~K and 200~K. The symbols of the lower
figure are obtained from the
data between 30~K and 110~K. The parameter $a = -0.0013~$K$^{-1}$ and $\rho_0 = 4.7~$m$\Omega$m.}
\label{2}
\end{figure}

\newpage

\begin{figure}
\centerline{\psfig{file=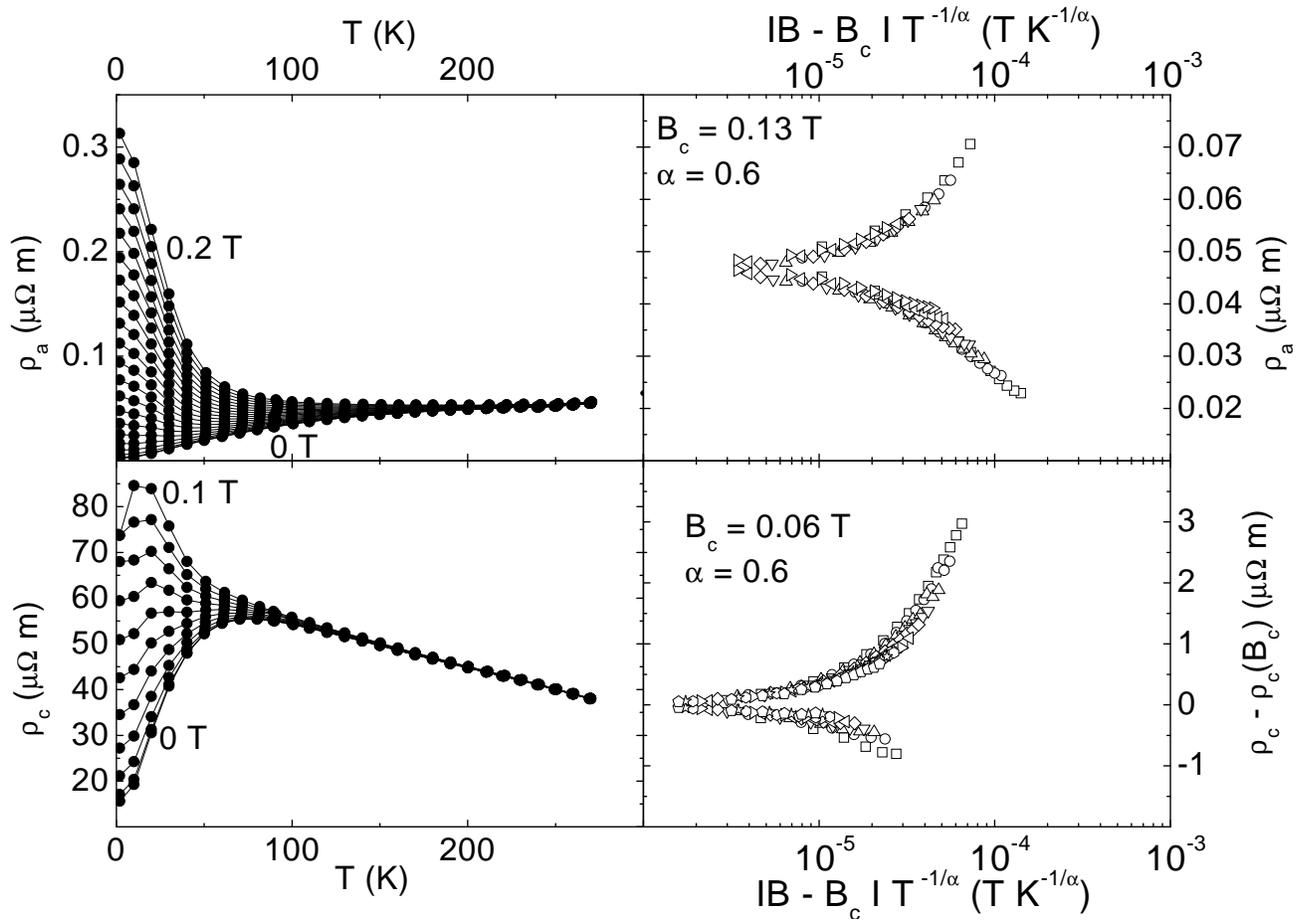,height=4.8in}}
\caption{The same as Fig. 1 for sample 3, $\rho_a(2$K$,0) = 0.0015 \mu\Omega$m. 
Right pannel: 
The symbols are the data points obtained between 50~K and 130~K. The symbols of the lower
figure correspond to the data taken between 90~K and 200~K.}
\label{3}
\end{figure}

\newpage

\begin{figure}
\centerline{\psfig{file=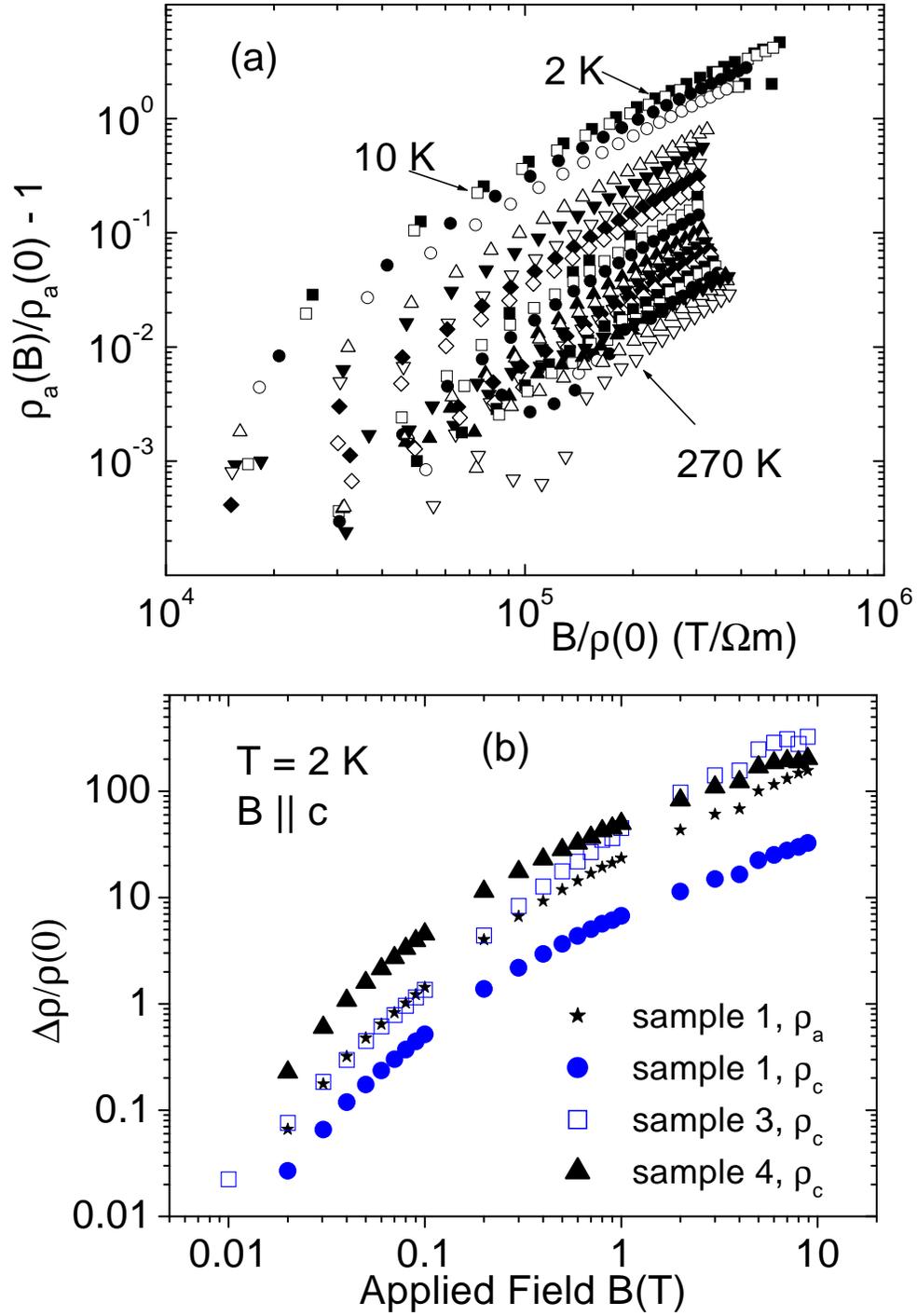,height=8.0in}}
\caption{(a) Kohler's plot of the in-plane magnetoresistance data for sample 1. Between 10~K
and 270~K each curve has been taken every 10~K. (b) Magnetoresistance $\Delta \rho = \rho_{a,c}(B) - \rho_{a,c}(0) /
\rho_{a,c}(0)$ as a function of applied field at $T = 2~$K for samples 1, 3 and 4.}
\label{4}
\end{figure}

\end{widetext}
\end{document}